\documentclass[pra,aps,twocolumn,showpacs,amsmath,amssymb]{revtex4-1}

\usepackage{graphicx}

\newcommand{\ket}[1]{|#1\rangle}
\newcommand{\bra}[1]{\langle #1|}
\newcommand{\braket}[2]{\langle #1|#2 \rangle}

\def\R{\textrm{I\kern-0.21emR}} 

\newcommand{\be}{\begin{equation}}
\newcommand{\ee}{\end{equation}}

\begin{document}

\title{Quantum versus classical effects in two-photon speckle patterns}

\author{Manutea Cand\'{e}}
\email{Manutea.Cande@grenoble.cnrs.fr}
\affiliation{Universit\'{e} Grenoble 1/CNRS, Laboratoire de Physique et Mod\'elisation des
Milieux Condens\'es UMR 5493,\\ B.P. 166, 38042 Grenoble, France}

\author{Sergey E. Skipetrov}
\email{Sergey.Skipetrov@grenoble.cnrs.fr}
\affiliation{Universit\'{e} Grenoble 1/CNRS, Laboratoire de Physique et Mod\'elisation des
Milieux Condens\'es UMR 5493,\\ B.P. 166, 38042 Grenoble, France}

\date{\today}

\begin{abstract}
We discuss quantum and classical aspects of two-photon interference in light transmission through disordered media. We show that disorder is the main factor that suppresses the interference, whatever the quantum state of the incident light. Secondarily, the two-photon interference is affected by the quantum nature of light (i.e., the well-defined number of photons in the two-photon entangled and Fock states as compared to the coherent state). And finally, entanglement is a resource that allows to prepare two-photon states with special symmetries with respect to the interchange of the photons and, in particular, the states with bosonic and fermionic symmetries. The two-photon interference is more robust for the latter states and its sign can be inverted for the fermionic state.
\end{abstract}


\maketitle

\section{Introduction}
\label{sec:intro}

Multiple scattering of light in random, disordered media is a subject of great fundamental and applied importance. Most of the studies in this field focus on classical phenomena that do not require quantization of the electromagnetic field to be understood \cite{Akkermans07, Vanrossum99, VanTiggelen03}. Light is even commonly categorized as a ``classical wave'' \cite{Vanrossum99}, in line with sound and elastic waves, as opposed to ``quantum waves'' describing electrons in disordered solids and matter waves in cold atomic systems. Meanwhile, even if states of light exist that mimic classical behavior very closely, light is probably the ``most quantum'' of all waves because it provides an unprecedented freedom in controlling and measuring its quantum state \cite{Nphoton09}. Indeed, the states of light that do not allow for a classical description (e.g., single and entangled photons, squeezed light, etc.) can nowadays be generated almost at will \cite{Kwiat95,Shields07}, opening very promising perspectives for communication and information processing applications, including quantum computation \cite{Nielsen10}.

Multiple scattering of non-classical light in random media attracts the attention of physicists since the pioneering paper by Beenakker \cite{Beenakker98} who introduced a convenient formalism of input-output relations in this field. In particular, propagation of pairs of entangled photons in random media has been studied both theoretically \cite{Beenakker09,Cherroret11} and experimentally \cite{Peeters10,Exter12} in recent years. Experimentally, entanglement can be studied and characterized by coincidence measurements: one measures the probability $P_2$ that two detectors each detect a photon simultaneously \cite{Hong87,Mandel95}. When two photons pass through some (simple) optical system (e.g., a beam-splitter, a delay line, or both in sequence), $P_2$ is affected by quantum interferences between different propagation paths leading to the same measurement outcome. Entanglement of photon states modifies the result of these interferences and can be read from the dependence of $P_2$ on the parameters of the optical system. Replacing the simple optical system by a random medium and assuming that the coincidence rate $P_2$ is measured as a function of the positions $\mathbf{r}_1$ and $\mathbf{r}_2$ of the two detectors, we arrive at the concept of the two-photon speckle pattern $P_2(\mathbf{r}_1, \mathbf{r}_2)$ \cite{Beenakker09}. This concept can be generalized to non-stationary (e.g., pulsed) light: the two-photon speckle pattern becomes time-dependent $P_2 = P_2(\mathbf{r}_1, t_1; \mathbf{r}_2, t_2)$. For two independent photons, $P_2$ factorizes: $P_2(1, 2) = P_1(1) P_1 (2)$, where we abbreviate $i = \{ \mathbf{r}_i, t_i \}$, $i = 1,2$, and $P_1(i)$ is the probability to detect a photon at a position $\mathbf{r}_i$ at a time $t_i$. It describes the usual, one-photon speckle pattern and it is proportional to the intensity of light at $\mathbf{r}_i$.

It is important to realize that $P_2$ defined above corresponds to a single realization of the random medium. It is therefore a random quantity and fluctuates from one realization of disorder to another. To obtain a deterministic quantity, it is therefore natural to average $P_2$ over an ensemble of realizations of the random medium \footnote{Higher-order statistical moments of $P_2$ were studied in Ref.\ \cite{Beenakker09}; its correlation functions might also be of interest.}. This ensemble-averaged quantity was studied in Refs.\ \cite{Beenakker09,Cherroret11,Peeters10,Exter12}. We will denote the ensemble average over random realizations of disorder in the medium by a horizontal bar $\overline{\vphantom{x}\cdots}$. It should be distinguished from the quantum-mechanical expectation value denoted by $\langle \cdots \rangle$. Note that even for two independent photons, $\overline{P_2(1,2)} = \overline{P_1(1) P_1(2)}$ does not factorize into a product of $\overline{P_1}$'s (in contrast to the unaveraged $P_2$) because $P_1(i)$ can have nontrivial (classical) correlations in both space and time \cite{Akkermans07,Vanrossum99}. Therefore, the ensemble-averaged two-photon speckle $\overline{P_2(1, 2)}$ combines properties due to the quantum nature of the incident light and those arising from the classical correlations between photons at two different positions (or times). This observation is the starting point of the analysis that we develop in the present paper. In the attempt to separate quantum effects from classical ones, we find that properties of two-photon speckles are conditioned by four distinct phenomena. First, the indistinguishability of photons plays a very important role. It does not require entanglement and can be present in both quantum non-entangled states (e.g., the two-photon Fock state) and the classical (i.e., coherent) state. Second, the quantum nature of the incident light is important. The quantumness of two-photon states, either entangled or not, is expressed by the fact that their number of photons ($n = 2$) is known and conserved. In contrast, the coherent state does not have a well defined number of photons associated to it. And third, the classical correlations between photons, induced by the fact that they propagate through the same random medium, turn out to be the most important factor. They determine the overall behavior of $\overline{P_2(1, 2)}$ whatever the state of the incident light. As a result, $\overline{P_2}$ calculated for two-photon entangled and non-entangled states, as well as for the classical light (coherent state) look very similar, with, however, one notable exception. Entanglement allows us to modify the symmetry of the state with respect to the exchange of the two photons \cite{Branning99,Exter12}. Results that are qualitatively different from those for Fock and coherent states are obtained for the antisymmetric state that changes sign upon this exchange. Photons behave as if they were fermions in this case. This behavior cannot be reproduced with non-entangled states and can be thus used as a hallmark of entanglement.

\section{The model}
\label{sec:model}

\subsection{Random medium and input-output relations}
\label{sec:disorder}

\begin{figure}
\centering
\includegraphics[width=8cm]{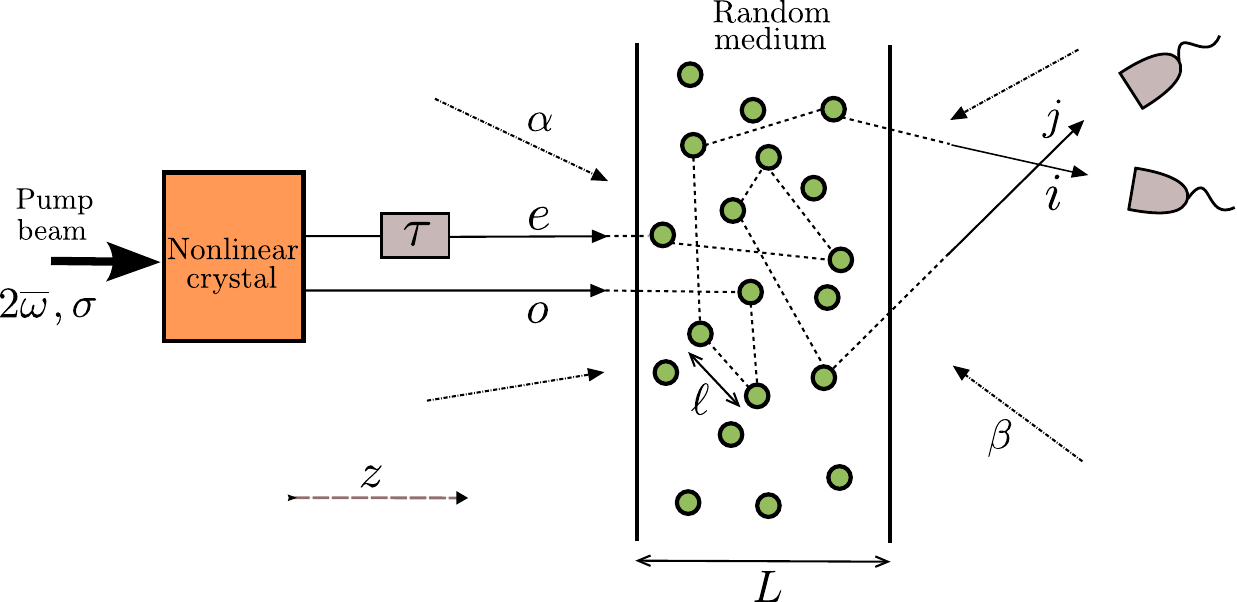}
\caption{Sketch of the experimental situation.
A birefrigent nonlinear crystal is pumped by laser pulses of central frequency $2 {\bar \omega}$ and bandwidth $\sigma$. Two collinear beams of orthogonal polarizations `o' (ordinary) and `e' (extraordinary) and frequencies $\omega_1$ and $\omega_2$, such that $\omega_1 + \omega_2 = 2 {\bar \omega}$, are obtained as a result of spontaneous parametric down-conversion in the crystal. These beams are incident on a slab of disordered medium, with the `e' beam delayed by a time $\tau$. Light is multiply scattered inside the slab and the transmitted light in modes $i$ and $j$ is detected by two photodetectors in the far field. The multiple scattering regime in ensured by the requirement $\ell \ll L$, with $\ell$ the mean free path due to disorder and $L$ the thickness of the slab.}
\label{fig:sketch}
\end{figure}

Consider a three-dimensional slab of elastically scattering random medium (no absorption or gain), perpendicular to the $z$ axis and having thickness $L$ and cross-section $A \gg L^2$ (see Fig.~\ref{fig:sketch}). The incident light is multiply scattered by the random heterogeneities of the medium before reaching one of the two detectors located in the far field. In the multiple scattering regime, the mean free path $\ell$ due to disorder is much less than the slab thickness $L$, $\ell \ll L$. On the other hand, the disorder is considered to be weak, so that $\ell \gg \lambda_0$, with $\lambda_0$ being the central wavelength of the incoming light. This corresponds to the diffuse regime in multiple scattering \cite{Akkermans07}.

Monochromatic light at a frequency $\omega$ incident on the slab can be decomposed over the basis of plane waves having wave vectors $\mathbf{k}(\omega) = \{ \mathbf{k}_{\perp}, k_z \}$, with $\mathbf{k}_{\perp} = \{ k_x, k_y \}$, in one of two orthogonal polarization states that we will denote by `o' and `e'. This notation refers to ordinary and extraordinary polarized waves in a birefrigent nonlinear crystal where, as we will explain below, the two waves can be generated (see Fig.\ \ref{fig:sketch}). For a slab of surface $A$, the number of modes in this basis is $2 \times N(\omega) = 2 \times k(\omega)^{2}A/2\pi$. The same representation is valid for the light leaving the slab.
From here on we will assume that $N$ can be assumed constant (i.e., independent of $\omega$) within the frequency bandwidth of the incident light. Input-output relations give the photon annihilation operators ${\hat a}_i(\omega)$ associated with the outgoing modes $i$ in terms of the annihilation operators ${\hat a}_{\alpha}(\omega)$ associated with the incoming modes $\alpha$ \cite{Beenakker98}:
\begin{equation}
\label{eq:input-output}
\hat{a}_{i }\left(\omega\right)=\sum_{\alpha=1}^{N}t_{\alpha i}\left(\omega\right)\hat{a}_{\alpha}\left(\omega\right)+
\sum_{\beta=N+1}^{2N}r_{\beta i}\left(\omega\right)\hat{a}_{\beta}\left(\omega\right).
\end{equation}
For the outgoing mode $i$ on the right of the slab (as in Fig.\ \ref{fig:sketch}), the first sum runs over the incoming modes $\alpha$ incident from the left and transmitted through the slab with transmission coefficients $t_{\alpha i}(\omega)$, whereas the second sum runs over the incoming modes $\beta$ incident from the right and reflected with reflection coefficients $r_{\beta i}(\omega)$. The coefficients $t_{\alpha i}(\omega)$ and $r_{\beta i}(\omega)$ form the (unitary) scattering matrix $S$ of the slab. Operators ${\hat a}_i$ and ${\hat a}_{\alpha}$ obey the usual bosonic commutation relations:
$[{\hat a}_{i}(\omega_{1}), {\hat a}_{j}(\omega_{2})] = 0$ and  $[{\hat a}_{i}(\omega_{1}), {\hat a}_{j}(\omega_{2})^{\dagger}] = \delta_{ij} \delta(\omega_{1}-\omega_{2})$, and the same for ${\hat a}_{\alpha}$ and ${\hat a}_{\beta}$.

As far as $L \gg \ell$, $N \gg 1$, and we are not interested in quantities that involve summations over all input or output modes (like, e.g., the conductance or the total transmission), the unitarity constraint on the matrix $S$ can be relaxed. Then, in the diffuse regime of scattering, the transmission coefficients $t_{\alpha i}(\omega)$ can be assumed to be independent identically distributed random variables with circular Gaussian statistics, zero mean, and some frequency correlation function $C(\Delta\omega)$: $\overline{t_{\alpha i}(\omega) t_{\alpha' j}(\omega')^*} = \overline{T} \delta_{\alpha \alpha'}
\delta_{ij} C(\omega-\omega')$.
Here $\overline{T} = \overline{|t_{\alpha i}(\omega) |^2}$ is the frequency-independent, average intensity transmission coefficient.

\subsection{Quantum states of light}
\label{sec:light}

The focus of this paper is on the two-photon entangled state that can be experimentally obtained by the spontaneous parametric down-conversion (SPDC) in a nonlinear optical crystal \cite{Klyshko88,Mandel95}. To be specific, we will restrict our consideration to the collinear type-II SPDC process in which two entangled photons have orthogonal polarizations (`o' and `e') and propagate collinearly with the pump beam \cite{Grice97}. At low intensity of the pump, the two-photon state generated in the SPDC process can be written as \cite{Grice97}
\begin{eqnarray}
\label{eq:entangledstate}
\ket{\psi_{\mathrm{ent}}} &=& \iint_{-\infty}^{\infty}d\omega_{1}d\omega_{2}B\left(\omega_{1},\omega_{2}
\right)\hat{a}_{\mathrm{o}}^{\dagger}\left(\omega_{1}\right)
\hat{a}_{\mathrm{e}}^{\dagger}\left(\omega_{2}\right)
\ket{0},\;\;\;
\end{eqnarray}
where $B\left(\omega_{1},\omega_{2}\right) =
K \alpha \left(\omega_{1}+\omega_{2}\right)
\Phi\left(\omega_{1},\omega_{2}\right)$ \footnote{Strictly speaking, the vacuum state $\left.|0\right>$ should be added to Eq.\ (\ref{eq:entangledstate}), but it would not contribute to any of results below, so that to lighten the notation, we omit it from the beginning.},
\begin{eqnarray}
\label{eq:alpha}
\alpha\left(\omega_{1}+\omega_{2}\right) &=& \exp[-(\omega_{1}+\omega_{2}-2\overline{\omega})^{2}
/2\sigma^{2}]
\end{eqnarray}
is the pump envelope and
\begin{eqnarray}
\label{eq:phi}
\Phi\left(\omega_{1},\omega_{2}\right) &=& \text{sinc}\left[\nu_{\mathrm{o}}
\left(\omega_{1}-\overline{\omega}\right)+
\nu_{\mathrm{e}}\left(\omega_{2}-
\overline{\omega}\right)\right]
\end{eqnarray}
follows from the phase matching condition \cite{Grice97}.
Here $2\overline{\omega}$ is the frequency and $\sigma$ is the spectral width of the pump beam;
$\nu_{j} = L_{\mathrm{NL}} \left(\partial k_{j}/\partial\omega|_{\omega = \overline{\omega}} - \partial k_{\mathrm{p}}/\partial\omega|_{\omega=2\overline{\omega}}\right)/2$, with $j = \mathrm{o}, \mathrm{e}$, quantify the phase mismatch between the down-converted photons `o' and `e' and the pump `p' in the nonlinear crystal of length $L_{\mathrm{NL}}$. $k_j(\omega)$ denotes the dispersion relation of the crystal for ordinary ($j = \mathrm{o}$) and extraordinary ($j = \mathrm{e}$) polarized light, as well as for the pump ($j = \mathrm{p}$).
The constant $K$ follows from the normalization condition
$\braket{\psi_{\mathrm{ent}}}{\psi_{\mathrm{ent}}} = |K|^2 \pi^{3/2} \sigma/|\eta_{-}| = 1$.

\begin{figure}
\centering
\includegraphics[width=8cm]{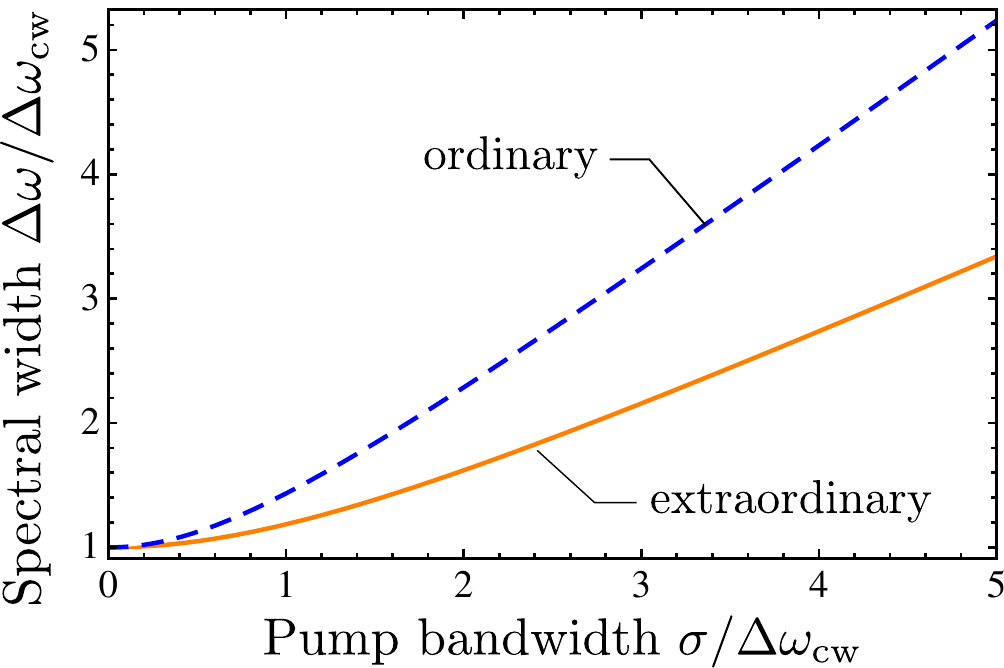}
\caption{Spectral widths of the ordinary and extraordinary beams of photons in the state described by Eq.\ (\ref{eq:entangledstate}), as functions of the bandwidth of the pump beam $\sigma$, for typical parameters of a BBO nonlinear crystal \cite{Grice01}. All quantities are normalized by the spectral width $\Delta \omega_{\mathrm{cw}}$ of the two beams in the monochromatic limit $\sigma = 0$.}
\label{fig:spectra}
\end{figure}

The entangled nature of the state (\ref{eq:entangledstate}) is manifest in the fact that the function $B(\omega_{1}, \omega_{2})$ is not factorizable in a product of two functions of $\omega_1$ and $\omega_2$, respectively. The two photons in the state (\ref{eq:entangledstate}) are thus entangled in frequency. A state similar to (\ref{eq:entangledstate}) was recently considered in Ref.\ \cite{Cherroret11}. In contrast to that work, however, our Eq.\ (\ref{eq:entangledstate}) describes a state that is not symmetric with respect to the exchange of frequencies of ordinary and extraordinary photons: $B(\omega_1, \omega_2) \ne B(\omega_2, \omega_1)$. This property stems from the different dispersion relations for the two polarization states in the nonlinear crystal and can, as we will see, have important consequences for the probability $P_2$ of simultaneous photon detection by two photodetectors because the two photons become more and more distinguishable as the bandwidth of the pump pulse $\sigma$ is increased. We illustrate this in Fig.\ \ref{fig:spectra} where we show the spectral width $\Delta \omega$, defined as the full width of the spectrum at half-maximum height, of ordinary and extraordinary beams, normalized by their value $\Delta \omega_{\mathrm{cw}} = 2.78/|\eta_{-}|$ in the monochromatic limit $\sigma \to 0$. To be able to estimate this quantity analytically, we replace the square of the $\mathrm{sinc}$ function in Eq.\ (\ref{eq:phi}) by a Gaussian $\exp(-x^{2}/2.79)$ with the same width at half-maximum. We obtain
\begin{eqnarray}
\label{eq:spectra}
\frac{\Delta\omega_{\mathrm{o, e}}}{\Delta\omega_{\mathrm{cw}}} = \frac{2|\nu_{\mathrm{o, e}}|}{|\eta_{-}|}\sqrt{\ln(2)
\left[ \frac{2.79}{\left(\nu_{\mathrm{o,e}}
\Delta\omega_{\mathrm{cw}}\right)^{2}} + \left(\frac{\sigma}{\Delta\omega_{\mathrm{cw}}}\right)^{2}\right]},
\;\;\;\;
\end{eqnarray}
where $\eta_{-}=\nu_{\mathrm{o}}-\nu_{\mathrm{e}}$ measures the typical delay between the two entangled photons when they leave the nonlinear crystal where they are generated. If the pump is monochromatic, $|\eta_{-}|$ can be considered as the coherence time of the `biphoton' \cite{Klyshko88}. We see from Eq.\ (\ref{eq:spectra}) and Fig.\ \ref{fig:spectra} that the spectral widths of the two beams become different when $\sigma$ increases, making distinguishable the photons originating from different beams after transmission through a disordered medium.

In order to understand the role of entanglement in the final results for $\overline{P}_2$, we will compare them to calculations performed for non-entangled states of light. The first non-entangled state that we will consider is the two-photon Fock state $\ket{\psi_{\mathrm{Fock}}}$ given by Eq.\ (\ref{eq:entangledstate}) with factorizable $B(\omega_1, \omega_2) = {\tilde B}(\omega_1) {\tilde B}(\omega_2)$. We will consider ${\tilde B} (\omega) =  \exp[-(\omega-\overline{\omega})^{2}/2\Delta^{2}]/(\sqrt{\pi} \Delta)^{1/2}$, where $\Delta$ is the bandwidth. This state has the same number of photons $n = 2$ as the entangled state (\ref{eq:entangledstate}) but does not imply any quantum correlations between them. However, the two photons have the same spectrum, so that they are indistinguishable after transmission through a disordered medium.

Finally, the simplest and the most widespread non-entangled state is the two-mode coherent state
\begin{equation}
\label{eq:coherentstate}
\ket{\psi_\mathrm{coh}} = \ket{\left\{\alpha\left(\omega\right)\right\}}_{\mathrm{o}}
\ket{\left\{\alpha\left(\omega\right)\right\}}_{\mathrm{e}},
\end{equation}
with $\alpha\left(\omega\right) = \exp[-(\omega-\overline{\omega})^{2}/
2\Delta^{2}]/(\sqrt{\pi} \Delta)^{1/2}$ \cite{Blow90}. It is an eigenstate of the photon annihilation operators ${\hat a}_{\mathrm{o}}(\omega)$ and ${\hat a}_{\mathrm{e}}(\omega)$: ${\hat a}_{\mathrm{o, e}}(\omega) \ket{\psi_\mathrm{coh}} = \alpha\left(\omega\right) \ket{\psi_\mathrm{coh}}$. This state describes two pulses with orthogonal polarizations. The fundamental difference between the coherent state (\ref{eq:coherentstate}) and the previously introduced two-photon states is that the former is not an eigenstate of the photon number operator ${\hat n}$. The number of photons is therefore not a good quantum number in this state. However, similarly to the Fock state, the photons in the state (\ref{eq:coherentstate}) have the same spectral properties and hence cannot be distinguished after being transmitted through a disordered medium.

\section{Average photocount rate}
\label{sec:intensity}

The entangled, Fock and coherent states introduced above have (on average for the coherent state) one photon in each polarization state. This can be confirmed by calculating the quantum expectation value $\langle {\hat n}_{\mathrm{o}, \mathrm{e}} \rangle$ of the photon number operator  \cite{Blow90}
\begin{eqnarray}
{\hat n}_{\mathrm{o}, \mathrm{e}} =
\int_{-\Delta T/2}^{\Delta T/2} dt\; {\hat a}_{\mathrm{o}, \mathrm{e}}^{\dagger}(t) {\hat a}_{\mathrm{o}, \mathrm{e}}(t),
\end{eqnarray}
where ${\hat a}_i(t)$ is the inverse Fourier transform of ${\hat a}_i(\omega)$. This quantity can be measured in an experiment by counting photons arriving at a polarization-discriminating photodetector placed in front of the photon source during a sufficiently long time $\Delta T$.

To start with and to illustrate our general calculation scheme, let us compute the average number of photons transmitted into a mode $i$ when the random medium is illuminated by light in one of the three states $\ket{\psi_{\mathrm{ent}}}$, $\ket{\psi_{\mathrm{Fock}}}$ or $\ket{\psi_{\mathrm{coh}}}$  defined in the previous section. The operator corresponding to this quantity is \cite{Blow90}
\begin{eqnarray}
\hat{n}_{i} = \int_{-\Delta T/2}^{\Delta T/2} dt\;  \hat{a}_{i}^{\dagger}(t) \hat{a}_{i}(t).
\label{eq:averagen}
\end{eqnarray}
To perform the calculation, we express the operator $\hat{a}_{i}(t)$ through the operators ${\hat a}_{\alpha}(t)$ and ${\hat a}_{\beta}(t)$ using the input-output relations (\ref{eq:input-output}), use commutation relations to compute the quantum expectation values and the Gaussian statistics of $t_{\alpha i}$ to average over realizations of disorder. The result is the same for the three quantum states that we study:
\begin{eqnarray}
\label{eq:intensity1}
\overline{\langle \hat{n}_{i} \rangle } =
(\langle {\hat n}_\mathrm{o} \rangle + \langle {\hat n}_\mathrm{e} \rangle) \overline{T} = 2 \overline{T}.
\end{eqnarray}
We thus observe that the average photon number is not sensitive to the quantum state of light, --- a well-known and quite general fact in quantum optics.

Given the normalization of the input states and the fact that $\overline{T} \ll 1$ in the diffuse regime, $\overline{\langle \hat{n}_{i} \rangle}$ is equal to the ensemble-averaged probability $\overline{P_1}$ that a photodetector measures a photon in the outgoing mode $i$. It also gives the photon counting rate (in units of photons per pulse) if a sequence of independent and separated in time pulses in quantum states $\ket{\psi_{\mathrm{ent}}}$, $\ket{ \psi_{\mathrm{Fock}}}$, or $\ket{\psi_{\mathrm{coh}}}$ is sent through the random medium.

\section{Coincidence rate and photon correlations}
\label{sec:coincidence}

Two different but related quantities can be used to characterize correlations between numbers of photons transmitted into two modes $i$ and $j$: the correlation function $C_{ij} = \langle {\hat n}_i {\hat n}_j \rangle$ and the probability $P_2(i, j)$ to detect a photon in each of the modes $i$ and $j$. The later also corresponds to the rate of photon coincidence counts if a sequence of pulses in the same quantum state is sent into the medium. The two quantities are related through a relation that we derive in Appendix \ref{app:corrvsrate}:
\begin{eqnarray}
P_2(i,j) = \frac{1}{1+\delta_{ij}} \langle:{\hat n}_i {\hat n}_j:\rangle = \frac{C_{ij} - \delta_{ij} \langle {\hat n}_i \rangle}{1+\delta_{ij}},
\label{eq:ratecorr}
\end{eqnarray}
where $: \cdots :$ denote the normal ordering of operators. It is worthwhile to note that $P_2(i,j)$ was identified with $\langle:{\hat n}_i {\hat n}_j:\rangle$ in some of the recent literature \cite{Cherroret11} without distinguishing the cases $i \ne j$ and $i = j$.

As we see from Eq.\ (\ref{eq:ratecorr}), both $P_2(i,j)$ and $C_{ij}$ can be found from the normally ordered correlation function $\langle:~{\hat n}_i {\hat n}_j~:\rangle$. The average of the latter over disorder can be computed for the three quantum states that we defined in Sec.\ \ref{sec:light} (see Appendix \ref{app:corrcalculation} for details):
\begin{eqnarray}
&&\overline{\langle: \hat{n}_{i} \hat{n}_{j} :\rangle}_{\mathrm{ent}} = 2 \overline{T}^2 \left[1 +
\delta_{ij} \iint_{-\infty}^{\infty} d\omega_{1} d\omega_{2}
B^{*}(\omega_{1},\omega_{2}) \right.
\nonumber \\
&&\;\;\;\times \left.  B(\omega_{2},\omega_{1})
|C(\omega_{1}-\omega_{2})|^{2} e^{-i(\omega_{1}-\omega_{2}) \tau} \vphantom{\int} \right],
\label{eq:coinc_ent1}
\\
&&\overline{\langle: \hat{n}_{i} \hat{n}_{j} :\rangle}_{\mathrm{Fock}} = 2 \overline{T}^2 \left[1 +
\delta_{ij} \iint_{-\infty}^{\infty} d\omega_{1} d\omega_{2}
|{\tilde B}(\omega_{1})|^2 \right.
\nonumber \\
&&\;\;\;\times \left. |{\tilde B}(\omega_{2})|^2 |C(\omega_{1}-\omega_{2})|^{2} e^{-i(\omega_{1}-\omega_{2}) \tau} \vphantom{\int} \right],
\label{eq:coinc_sep1}
\\
&&\overline{\langle: \hat{n}_{i} \hat{n}_{j} :\rangle}_{\mathrm{coh}} =  4 \overline{T}^2
\left[1 + \delta_{ij} \iint_{-\infty}^{\infty} d\omega_{1} d\omega_{2} |\alpha(\omega_{1})|^{2} \right.
\nonumber \\
&&\;\;\;\times \left. |\alpha(\omega_{2})|^{2} |C(\omega_{1}-\omega_{2})|^{2}
\cos^2([\omega_{1}-\omega_{2}]\tau/2) \vphantom{\int}\right].
\label{eq:coinc_coh1}
\end{eqnarray}

In the multiple scattering regime, the quantities given by Eqs.\ (\ref{eq:coinc_ent1})--(\ref{eq:coinc_coh1}) will be very small and one can doubt their interest for real experiments. Indeed, typically we have $\overline{T} \sim \ell/NL$ with $L/\ell \gtrsim 10$. The number of transverse modes for a slab of area $A = 1$ mm$^2$ will be $N = k^2 A/2\pi \sim 10^7$ at optical frequencies, leading to a prefactor $\overline{T}^2 \sim 10^{-16}$ in Eqs.\ (\ref{eq:coinc_ent1})--(\ref{eq:coinc_coh1}). Nevertheless, photon coincidence measurements in transmission through a disordered medium were realized for coherent and quasi-chaotic states \cite{Smolka11}. The measurement can be optimized by noting that Eqs.\ (\ref{eq:coinc_ent1})--(\ref{eq:coinc_coh1}) do not depend on the particular choice of modes $i$ and $j$ but are only sensitive to the fact that $i = j$ or $i \ne j$. One can thus count photons in an arbitrary large number $M > 2$ of modes and then average over the results obtained for all pairs $(i, j)$ of modes. Each mode should, however, be addressed individually, which will require $M$ photodetectors. One can also think about more sophisticated experimental setups to detect photon coincidences without knowing the precise mode to which the photons belong, like the one exploiting the two-photon absorption \cite{Boitier09}. From the theoretical point of view, we can avoid working with too small quantities by normalizing Eqs.\ (\ref{eq:coinc_ent1})--(\ref{eq:coinc_coh1}) by ${\overline{T}}^2$. We thus define the normalized photocount coincidence rate $R$ as
\begin{eqnarray}
R_{ij} &=& \frac{1}{\overline{T}^2} \overline{P_2(i,j)}
= \frac{1}{\overline{T}^2} \times \frac{\overline{\langle: {\hat n}_i {\hat n}_j :\rangle}}{1+\delta_{ij}}.
\label{eq:rater}
\end{eqnarray}

An expression similar in structure to Eq.\ (\ref{eq:coinc_ent1}) was derived for the coincidence rate in Ref.\ \cite{Cherroret11}, though for a slightly different entangled state. In addition, the authors of Ref.\ \cite{Cherroret11} have taken into account the lowest-order corrections to the Gaussian model for transmission coefficients $t_{\alpha i}$ adopted here, and found additional contributions to the coincidence rate. These contributions, however, turned out to be of order $1/g$, with $g = N \ell/L \gg 1$, and would be negligible under conditions of diffuse scattering that we consider here. In the absence of these terms, we find from Eqs.\ (\ref{eq:coinc_ent1})--(\ref{eq:rater}) that for $i \ne j$, $R_{\mathrm{ent}} = R_{\mathrm{Fock}} = 2$ and $R_{\mathrm{coh}} = 4$ independent of any parameters. The latter result corresponds to the total absence of any correlation between photon numbers in two different modes $i \ne j$: $\langle: {\hat n}_i {\hat n}_j :\rangle = 4 \overline{T}^2 = \langle {\hat n}_i \rangle \langle {\hat n}_j \rangle$, whereas the former indicates that there are negative correlations: $\langle: {\hat n}_i {\hat n}_j :\rangle = 2 \overline{T}^2 < \langle {\hat n}_i \rangle \langle {\hat n}_j \rangle$. Negative correlations in transmission of non-classical light through disordered media were already predicted theoretically in Refs.\ \cite{Lodahl05,Skipetrov07,Ott10} and observed experimentally in Ref.\ \cite{Smolka09} for squeezed light.

For $i \ne j$, the parameter-dependent corrections to Eqs.\ (\ref{eq:coinc_ent1})--(\ref{eq:coinc_coh1}) due to the non-Gaussian statistics of $t_{\alpha i}$ and correlations between them would be of order $1/g \ll 1$ \cite{Cherroret11}. This is much smaller than the variations of the coincidence rates by 50\% or so for $i = j$. The latter should be therefore much easier accessible experimentally. For this reason, in the rest of this paper we will study the case $i = j$ and compute $R_{ii}$ which describes the probability for two photons to be found in the same outgoing mode. We will omit the subscript `$ii$' of $R_{ii}$ to lighten the notation. At this point, the lacking ingredient in Eqs.\ (\ref{eq:coinc_ent1}), (\ref{eq:coinc_sep1}) and (\ref{eq:coinc_coh1}) is the frequency correlation function $C(\Delta \omega)$. We will use two different models for it. A simple model I, $C(\Delta \omega) = \exp(-|\Delta \omega|/\Omega_{\mathrm{corr}})$, with $\Omega_{\mathrm{corr}}$ the correlation frequency, will facilitate integrations in Eqs.\ (\ref{eq:coinc_ent1})--(\ref{eq:coinc_coh1}) and thus will allow for a number of analytic results. A more realistic model II, $C(\Delta \omega) = \sqrt{-i \Delta \omega/\Omega_{\mathrm{Th}}}/\sinh(\sqrt{-i \Delta \omega/\Omega_{\mathrm{Th}}})$ \cite{Sebbah00}, with $\Omega_{\mathrm{Th}}$ the Thouless frequency, will be used to obtain predictions that can be experimentally verified in a diffusely scattering disordered medium. We note, however, that both correlation function given above decay fast with $\Delta \omega$ and hence the results will be similar for both models.

\subsection{Entangled state}
\label{sec:entangled}

For the entangled state (\ref{eq:entangledstate}) and the exponential model I for the correlation function $C(\Delta \omega)$ of transmission coefficients, we obtain from Eq.\ (\ref{eq:coinc_ent1}):
\begin{eqnarray}
\label{eq:coinc_ent2}
&&R_{\mathrm{ent}}(t,s,w)
\nonumber \\
&&\hspace*{5mm}= 1 + \frac{1}{s\sqrt{\pi}} \int_{-1}^{1}dx\; f\left(t,w,x\right)
\mathrm{Erf}\left[\frac{s}{2}\left(1-|x|\right)\right],\;\;\;
\end{eqnarray}
where $f(t,w,x)
=2w/[4+w^{2}\left(x+t\right)^{2}]$ and the dimensionless variables are $t = \tau/\eta_{-}$, $s = \sigma\eta_{+}$ and $w=|\Omega_{\mathrm{corr}}\eta_{-}|$, with $\eta_{\pm}=\nu_{\mathrm{o}} \pm \nu_{\mathrm{e}}$.

\begin{figure}
\centering
\includegraphics[width=8cm]{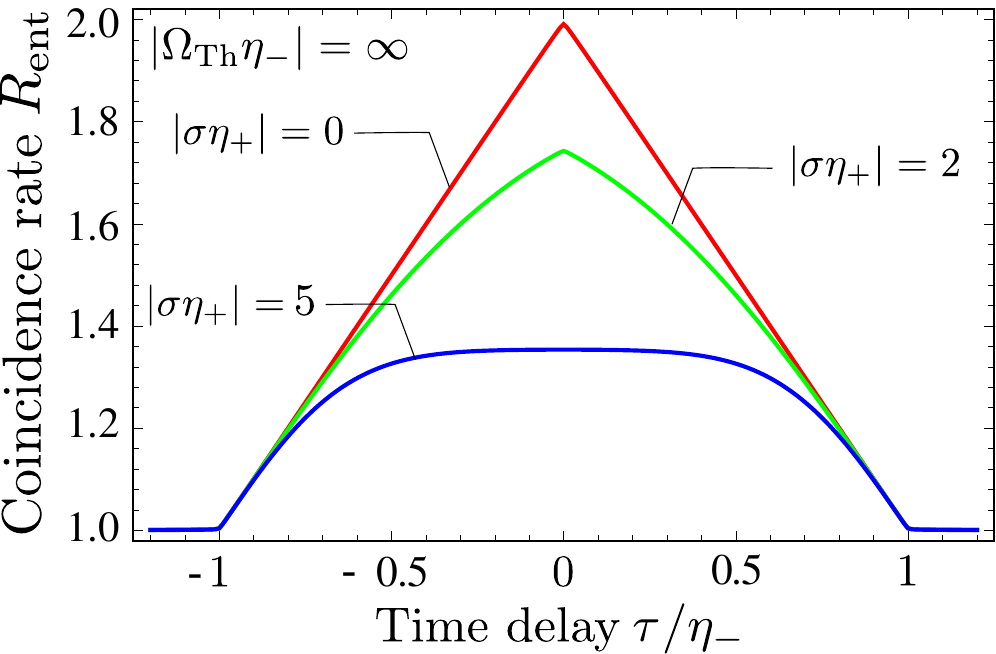}
\caption{Photocount coincidence rate for two detectors in the same outgoing mode and the two-photon entangled state (\ref{eq:entangledstate}) at the input of a disordered medium with frequency-independent transmission coefficients $t_{\alpha i}$, as a function of delay $\tau$ between the two photons. Three curves correspond to different values of the parameter $s = \sigma\eta_{+}$ measuring the spectral width of the pulse that pumps the nonlinear crystal where the photons are generated. An infinitely large value of Thouless frequency is assumed.}
\label{fig:hom}
\end{figure}

For $\Omega_{\mathrm{corr}} \gg 1/|\eta_{-}|$, $t_{\alpha i}(\omega)$ can be assumed frequency-independent within the bandwidth $1/|\eta_{-}|$ around $\bar{\omega}$ and Eq.\ (\ref{eq:coinc_ent2}) reduces to
\begin{eqnarray}
\label{eq:HOM}
&&R_{\mathrm{ent}} (t,s,\infty)
\nonumber \\
&&\hspace*{1cm}
= 1 + \frac{\sqrt{\pi}}{s}\left\{ \begin{array}{ll}\text{Erf}\left[\frac{s}{2}\left(1-|t|\right)\right] & ,|t| < 1\\
0  & ,|t| \geq 1\\
\end{array} \right.
\end{eqnarray}
Exactly the same result is obtained for the model II in the limit of $|\Omega_{\mathrm{Th}} \eta_{-}|\to \infty$.
Up to a sign in front of the second term in the r.h.s., Eq.\ (\ref{eq:HOM}) is equivalent to the coincidence rate in the Hong-Ou-Mandel interferometer \cite{Grice97}. Instead of a dip in the coincidence rate as a function of delay $\tau$, we now find a peak, which is explained by the fact that we compute coincidences in the same outgoing mode and not in two different modes. We show Eq.\ (\ref{eq:HOM}) in Fig.\ \ref{fig:hom} for 3 values of $s$, corresponding to 3 different bandwidths $\sigma$ of the pump pulse at fixed $\eta_+$. Increasing $\sigma$ suppresses the two-photon interference effect in $R_{\mathrm{ent}}$. This is due to the loss of indistinguishability of the two photons, which, in its turn, results from the fact that the two entangled photons of the state (\ref{eq:entangledstate}) do not have the same spectrum (see Fig.\ \ref{fig:spectra}). As a consequence, the two-photon interference between them washes out with increase of $\sigma$.

The minimum and the maximum values $R = 1$ and 2 of the normalized coincidence rate in Fig.\ \ref{fig:hom} can be understood without appealing to the entangled nature of the quantum state and turn out to be universal for all two-photon states. For $|\tau| > |\eta_{-}|$, the photons are well separated in time and are transmitted through the disordered medium independently. The probability for each photon to end up in a given outgoing mode $i$ is equal to $\overline{T}$. The probability that both photons are measured in the same mode $i$ is equal to the product of single-photon probabilities: $\overline{P_2(i,j)} = \overline{T}^2$, leading to $R = 1$. In contrast, at $\tau = 0$ the two photons cannot be considered independent and the two-photon interference leads to a larger probability for them to be detected in the same mode, thus the peak in Fig.\ \ref{fig:hom}. In the ideal case of $\sigma \eta_{+} = 0$ this interference is perfectly constructive and $R$ doubles with respect to the case of independent photons. We will show in Sec.\ \ref{sec:symmetry} that {\em destructive} two-photon interference can lead to $R < 1$ and even to $R = 0$ for certain specially prepared states.

\begin{figure}
\centering
\includegraphics[width=8cm]{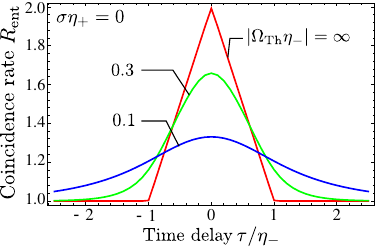}
\caption{
Photocount coincidence rate for two detectors in the same outgoing mode and the two-photon entangled state (\ref{eq:entangledstate}) at the input of a disordered medium. The pump beam that generates the pair of photons in a nonlinear crystal is assumed to be monochromatic ($\sigma = 0$); different values of $|\Omega_{\mathrm{Th}}\eta_{-}|$ correspond to different disorder strengths and/or sample thicknesses.
}
\label{fig:entdisorder}
\end{figure}

Let us now consider the case of arbitrary relation between the Thouless frequency $\Omega_{\mathrm{Th}}$ and $\eta_{-}$. In Fig.\ \ref{fig:entdisorder}, we plot the coincidence rate for 3 different values of $|\Omega_{\mathrm{Th}} \eta_{-}|$. We observe that the increase of disorder (quantified by the decrease of the parameter $|\Omega_{\mathrm{Th}} \eta_{-}|$) has two distinct consequences. First, we observe lowering of the maximum coincidence rate in the absence of delay ($\tau = 0$). The disorder thus washes out the two-photon interference. However, at the same time we see that the curve $R_{\mathrm{ent}}(\tau)$ broadens and the photon coincidence rate grows up at large $|\tau| \gtrsim |\eta_{-}|$. This is a consequence of the fluctuating time of flight of photons through the medium. Typically, the time of flight of a photon through a disordered medium fluctuates from very small values (ballistic propagation) to values exceeding $\Omega_{\mathrm{Th}}^{-1}$ (diffusion). The medium can, therefore, partially compensate for the initial delay $\tau$ between the two photons and make them interfere even for $|\tau| > |\eta_{-}|$.

\subsection{Separable state}\label{sec:separable}

As follows from Figs.\ \ref{fig:hom} and \ref{fig:entdisorder}, the photocount coincidence rate for two photodetectors behind a disordered medium is sensitive to the parameters $\sigma$ and $\eta_{-}$ that characterize the state of the two entangled photons incident on the medium. But let us check what really changes if now we consider the two-photon Fock state $\ket{\psi_{\mathrm{Fock}}}$ defined in Sec.\ \ref{sec:light}. This state is separable (not entangled), but the two photons are still indistinguishable. Therefore, the comparison of results obtained for the states $\ket{\psi_{\mathrm{ent}}}$ and $\ket{\psi_{\mathrm{Fock}}}$ should allow revealing the role of entanglement in the experiment depicted in Fig.\ \ref{fig:sketch}. For the exponentially decaying correlation function $C(\Delta \omega)$, we obtain from Eqs.\ (\ref{eq:ratecorr}) and (\ref{eq:coinc_sep1}):
\begin{eqnarray}
\label{eq:coinc_sep2}
R_{\mathrm{Fock}}(t, w) &=& 1 +
\mathrm{e}^{-\frac{1}{2} \left[|t|^{2}-\left(\frac{2}{w}\right)^{2}\right]}
\left\{\vphantom{\frac{s}{\sqrt{2}}} \mathrm{cos}\left(\frac{2|t|}{w}\right)\right.
 \nonumber\\
  &-& \left. \mathrm{Re} \left( \mathrm{e}^{-2i\frac{|t|}{w}}\mathrm{Erf}
  \left[\frac{1}{\sqrt{2}}\left(\frac{2}{w}- i|t|\right)\right]\right)\right\},
  \;\;\;\;\;\;
\end{eqnarray}
where $t = \tau\Delta$ and $w = \Omega_{\mathrm{corr}}/\Delta$.

\begin{figure}
  \centering
  \includegraphics[width=8cm]{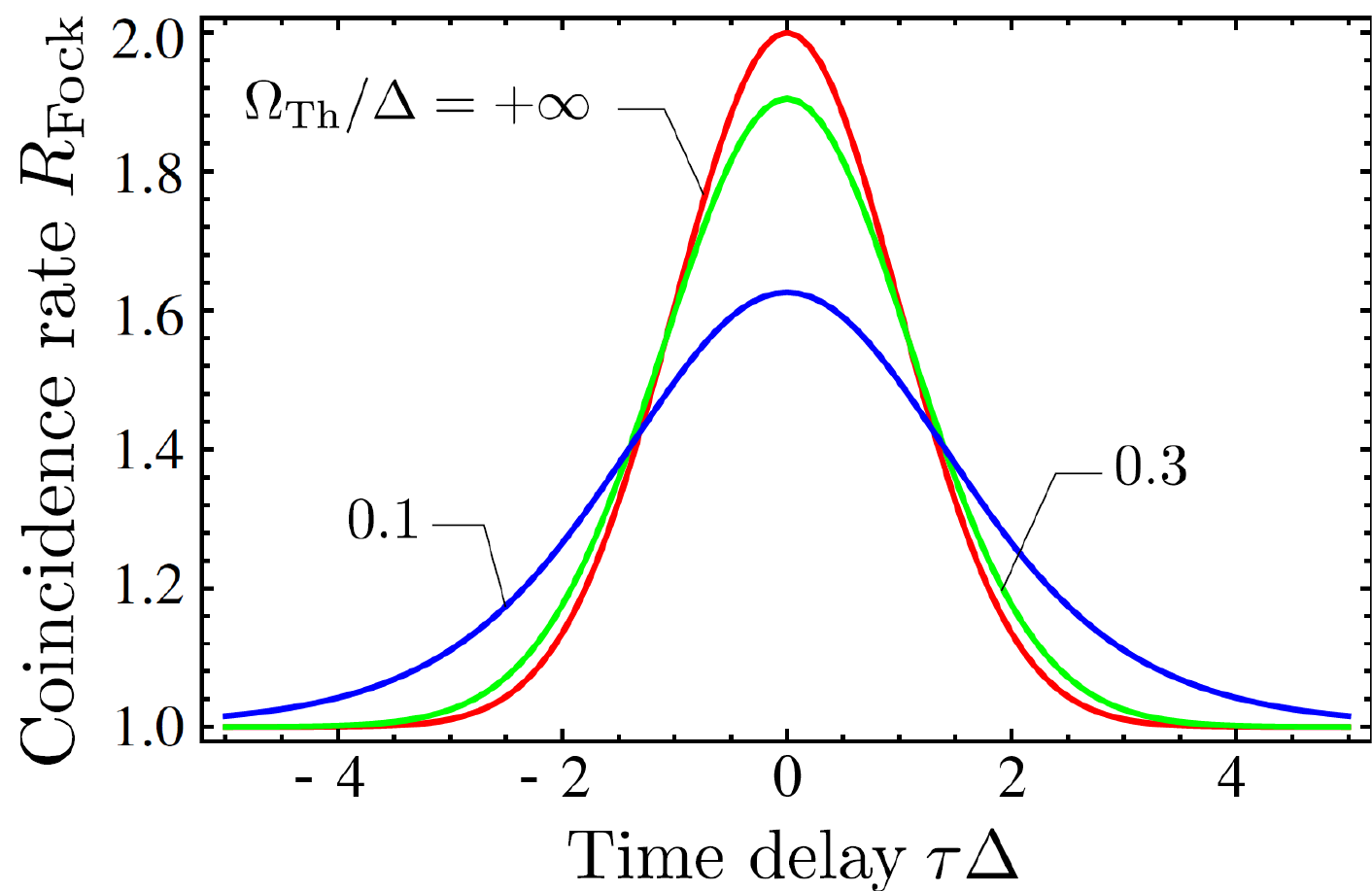}
  \caption{Same as Fig.\ \ref{fig:entdisorder} but for the separable, two-photon Fock state incident on the disordered medium. The delay time $\tau$ and the Thouless frequency $\Omega_{\mathrm{Th}}$ are in units of bandwidth $\Delta$ of incident photon beams.}
  \label{fig:sep1}
\end{figure}

In Fig.\ \ref{fig:sep1} we show the coincidence rate $R_{\mathrm{Fock}}$ for the more realistic model II. $R_{\mathrm{Fock}}$ is shown as a function of normalized time delay at three different values of the normalized Thouless frequency. It is bounded by 1 from below and 2 from above for the same reasons as $R_{\mathrm{ent}}$ (see the discussion in Sec.\ \ref{sec:entangled}). Figure \ref{fig:sep1} is to be compared with Fig.\ \ref{fig:entdisorder}.
At weak disorder (large $\Omega_{\mathrm{Th}}$) the results for the entangled and separable two-photon states are quite different: the triangular shape of $R_{\mathrm{ent}}(\tau)$ is replaced by a rounded, Gaussian-like profile of $R_{\mathrm{Fock}}(\tau)$. In contrast, at small $\Omega_{\mathrm{Th}}$ the two states lead to similar results. We thus can obtain very similar dependencies $R(\tau)$ for two-photon entangled and separable states, provided that we adjust the bandwidth of the latter to values $\Delta \sim 1/|\eta_{-}|$. Given this observation, we are led to conclude that, at least in the presence of sufficiently strong disorder (i.e., for $|\Omega_{\mathrm{Th}} \eta_{-}|$, $\Omega_{\mathrm{Th}}/\Delta \lesssim 1$), the impact of entanglement on the coincidence rate is quite limited. In an experiment measuring $R(\tau)$ behind a random medium, it would be difficult (if possible at all) to conclude if the incident state was the entangled state $\ket{\psi_{\mathrm{ent}}}$ or the separable state  $\ket{\psi_{\mathrm{Fock}}}$ when no additional information (e.g., the bandwidth of the incident photon pulses) is available.

\subsection{Coherent state}
\label{sec:coherent}

Because in transmission through a disordered medium the two-photon entangled and separable states lead to very similar dependencies of the coincidence rate on the delay time between the two photons, we would like to clarify the common property of these states responsible for this result. Clearly, both states $\ket{\psi_{\mathrm{ent}}}$ and $\ket{\psi_{\mathrm{Fock}}}$ contain exactly two photons and these two photons are indistinguishable. But what is more important? To answer this question, we now consider the coherent state $\ket{\psi_{\mathrm{coh}}}$ in which the number of photons is not well-defined, but the photons are still indistinguishable. Assuming the model I for the correlation function $C(\Delta \omega)$, we can again compute the coincidence rate analytically from Eqs.\ (\ref{eq:coinc_coh1}) and (\ref{eq:rater}):
\begin{eqnarray}
\label{eq:coinc_coh2}
R_{\mathrm{coh}}(t, w) &=&
2 +
\mathrm{e}^{2/w^{2}}\mathrm{Erfc}\left(\frac{\sqrt{2}}{w}\right) \nonumber\\
&+& \mathrm{e}^{-\frac{1}{2}\left[ |t|^{2}-\left(\frac{2}{w}\right)^{2}\right]}
\left\{\vphantom{\frac{s}{\sqrt{2}}} \mathrm{cos}\left(\frac{2|t|}{w}\right)\right.
 \nonumber\\
  &-& \left. \mathrm{Re} \left( \mathrm{e}^{-2i\frac{|t|}{w}}\mathrm{Erf}
  \left[\frac{1}{\sqrt{2}}\left(\frac{2}{w} - i|t|\right)\right]\right)\right\},
  \;\;\;\;\;\;
\end{eqnarray}
where $t =\tau\Delta$ and $w = \Omega_{\mathrm{corr}}/\Delta$.

\begin{figure}
  \centering
  \includegraphics[width=8cm]{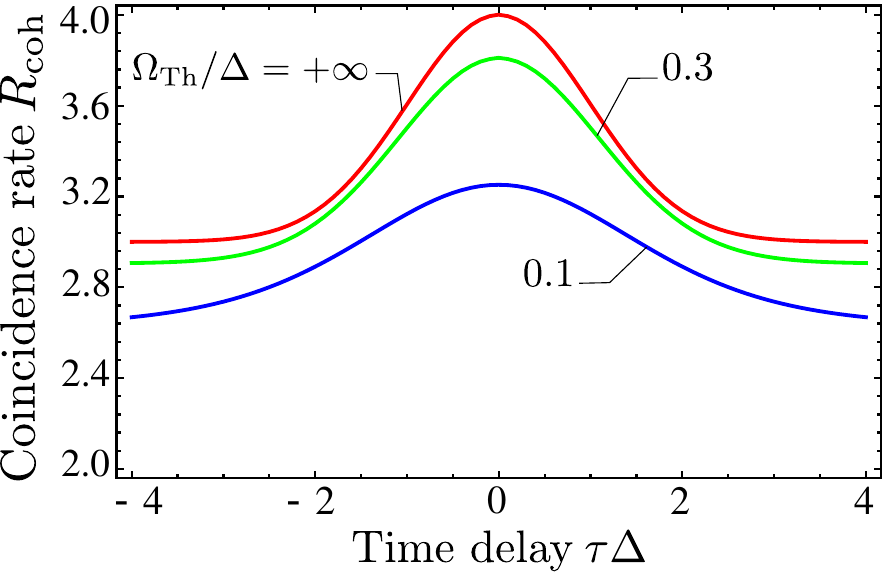}
  \caption{Same as Figs.\ \ref{fig:entdisorder} and \ref{fig:sep1} but for the coherent state (\ref{eq:coherentstate}) incident on the disordered medium. The delay time $\tau$ and the Thouless frequency $\Omega_{\mathrm{Th}}$ are in units of bandwidth $\Delta$.}
  \label{fig:coh1}
\end{figure}

In Fig.\ \ref{fig:coh1} we show $R_{\mathrm{coh}}$ as a function of delay time $\tau$ for the realistic model II of the correlation function $C(\Delta \omega)$. Despite the fact that the coherent state exhibits neither entanglement, nor a well-defined photon number, the overall shape of the coincidence curve is still the same: it is a bell-shaped curve with the overall amplitude suppressed by disorder.
However, $R_{\mathrm{coh}}$ reaches larger absolute values, up to $R_{\mathrm{coh}} = 4$, in contrast to $R_{\mathrm{ent}}$ and $R_{\mathrm{Fock}}$ that are bounded by 2 from above.

The results for the coherent state can be rederived in a completely classical framework, by replacing the operators ${\hat a}$ and ${\hat a^{\dagger}}$ with complex numbers $a$ and $a^*$ and ignoring the commutation relations. This provides a way to understand some of the aspects of Fig.\ \ref{fig:coh1} using the well-known facts from the theory of classical wave scattering \cite{Akkermans07,Vanrossum99}. Indeed, we obtain from Eq.\ (\ref{eq:rater}) that $R_{\mathrm{coh}} = 2 \overline{I_{\Delta T}^2}/(\overline{I_{\Delta T}})^2$, where $I_{\Delta T}$ is the intensity of the transmitted wave, integrated over a time interval $\Delta T$. When $\tau = 0$, the transmitted wave $a_i$ is a superposition of independent waves $a_i^{(\mathrm{o})}$ and $a_i^{(\mathrm{e})}$ resulting from the transmission of incident waves with `o' and `e' polarizations, respectively. For weak disorder, both $a_i^{(\mathrm{o})}$ and $a_i^{(\mathrm{e})}$ (and hence $a_i$) have zero-mean circular Gaussian statistics. Because the intensity $I = |a_i|^2$, $\overline{I_{\Delta T}^2}/(\overline{I_{\Delta T}})^2 = 2$ is obtained for the monochromatic illumination ($\Delta \rightarrow 0$). This corresponds to the so-called large intensity fluctuations, well-known for classical speckle patterns \cite{Shapiro86}. The resulting maximum value of $R_{\mathrm{coh}}$ is 4, as can also be seen from Fig.\ \ref{fig:coh1}. In contrast, when $|\tau|$ is large, the detections of signals due to the incident `o' and `e' waves are separated in time. The measured time-integrated intensity is a sum of two independent terms: $I_{\Delta T} = I_{\Delta T}^{(\mathrm{o})} + I_{\Delta T}^{(\mathrm{e})}$. This yields $\overline{I_{\Delta T}^2}/(\overline{I_{\Delta T}})^2 = 3/2$ and $R_{\mathrm{coh}} = 3$ for $\Delta \to 0$. When the spectral width $\Delta$ of the incident waves increases, the measured signals suffer from a partial averaging even for a single realization of disorder. In the limit of $\Delta \to \infty$, the intensity of transmitted light does not fluctuate and $R_{\mathrm{coh}} = 2$.


\section{Symmetric and antisymmetric entangled states}
\label{sec:symmetry}

As we have shown in the previous section, the entangled state (\ref{eq:entangledstate}) with the function $B(\omega_1, \omega_2)$ defined by Eqs.\ (\ref{eq:alpha}) and (\ref{eq:phi}) leads to the coincidence rate $R$ measured behind a random medium that is similar to that for a separable state (e.g., compare Figs.\ \ref{fig:entdisorder} and \ref{fig:sep1}). In an experiment, if one simply measures $R$ as a function of delay $\tau$, it might be difficult or even impossible to conclude on the presence of entanglement between the two incident photons without knowing their other characteristics, such as the bandwidth. This might give an impression that entanglement adds no interesting physics to the optics of random media. This impression is, however, completely wrong. In order to demonstrate this, let us now consider a little bit more sophisticated entangled state, namely, the state given once again by Eq.\ (\ref{eq:entangledstate}) where we replace $B(\omega_1, \omega_2)$ by
\begin{eqnarray}
B_{\theta}(\omega_{1}, \omega_{2}) = {\cal K} [B(\omega_{1},\omega_{2}) + \mathrm{e}^{i\theta}B(\omega_{2},\omega_{1})],
\label{eq:btheta}
\end{eqnarray}
with ${\cal K}$ being a normalization constant,
\begin{eqnarray}
|{\cal K}|^2 = \frac{1}{2} \times \frac{1}{1 + \cos(\theta)
\mathrm{Erf}(\sigma \eta_+/2) \sqrt{\pi}/\sigma \eta_+}.
\label{eq:symnorm}
\end{eqnarray}
By properly adjusting the value of $\theta$, we can make this state symmetric ($\theta = 0$) or antisymmetric ($\theta = \pi$) with respect to the exchange of the two photons. In other words, the state preserves its form with the same (symmetric) or opposite (antisymmetric) sign upon the exchange $\omega_1 \leftrightarrow \omega_2$ and can thus also be said to have bosonic or fermionic symmetry, respectively. An intermediate situation with $0 < \theta < \pi$ corresponds to asymmetric states; $\theta = \pi/2$, for example, yields a state that leads to the same result for the average two-photon coincidence rate as the state (\ref{eq:entangledstate}) that we studied in Sec.\ \ref{sec:entangled}.

The states corresponding to Eq.\ (\ref{eq:btheta}) can be prepared experimentally \cite{Branning99} and for $\theta = 0$, $\pi$ have the advantage of conserving the perfect indistinguishability of the two photons whatever the bandwidth of the pump $\sigma$. This is in contrast to the entangled state considered in the previous sections, in which the two photons become distinguishable as $\sigma$ is increased because their spectra become different (see Fig.\ \ref{fig:spectra}). As a result, Eq.\ (\ref{eq:btheta}) leads to a much weaker dependence of the coincidence rate on $\sigma$. In addition, the sign of the two-photon interference term can be inversed for the antisymmetric state ($\theta = \pi$), thus turning the constructive interference into the destructive one.

Because $B_{\theta}(\omega_1, \omega_2)$ is a linear combination of $B(\omega_1, \omega_2)$ and $B(\omega_2, \omega_1)$, the calculation of the coincidence rate for the former can be reduced to the analysis that was performed in Sec.\ \ref{sec:entangled}. For the exponentially decaying correlation function $C(\Delta \omega)$ we obtain
\begin{eqnarray}
\label{eq:coinc_sym}
R_{\theta}(t, s, w) &=& 1 + 2 \delta_{ij} |{\cal K}|^{2}
\int_{-1}^{1} dx f(t, w, x)
\nonumber \\
&\times& \left[ I(s, x) + \cos(\theta) J(s, x)\right],
\end{eqnarray}
where $f(t, w, x)$ was defined in Sec.\ \ref{sec:entangled}. The functions $I(s, x)$ and $J(s, x)$ are
\begin{eqnarray}
I(s,x) &=& \frac{1}{s\sqrt{\pi}}\mathrm{Erf}\left[\frac{s}{2}\left(1-
|x|\right)\right],\\
J(s,x) &=& \frac{1}{\pi}\left(1-|x|\right)
\exp\left(-\frac{s^{2}x^{2}}{4} \right).
\end{eqnarray}

\begin{figure}
\centering
\includegraphics[width=8cm]{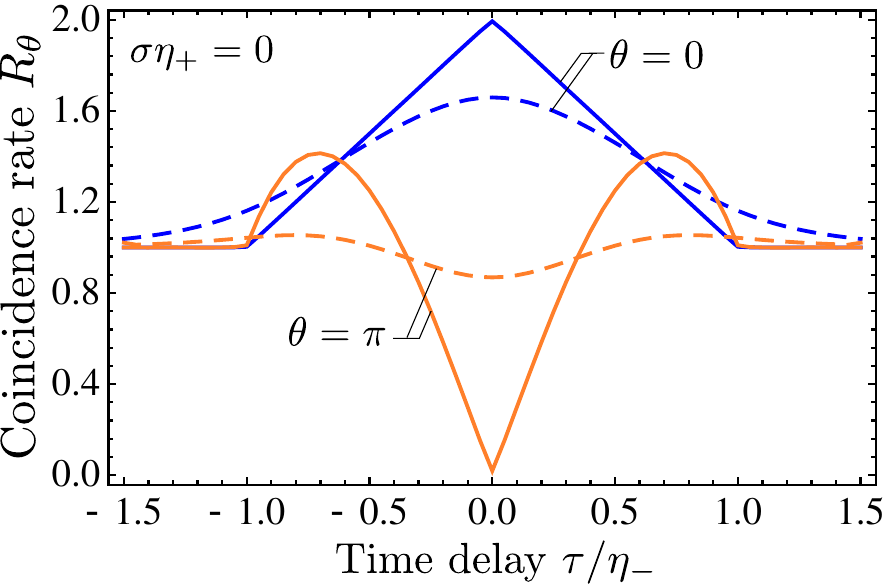}
\caption{Photocount coincidence rate for two detectors in the same outgoing mode and the symmetric ($\theta = 0$, blue lines) and antisymmetric ($\theta = \pi$, orange lines) entangled states at the input of a disordered medium. The pump is monochromatic and the disorder is weak ($|\Omega_{\mathrm{Th}} \eta_{-}| \to \infty$) for solid lines.
$|\Omega_{\mathrm{Th}} \eta_{-}| = 0.3$ for dashed lines.}
\label{fig:symantisym}
\end{figure}

The striking difference between symmetric ($\theta = 0$) and antisymmetric ($\theta = \pi$) states is demonstrated in Fig.\ \ref{fig:symantisym}. We see that for a monochromatic pump ($\sigma = 0$) and weak disorder ($|\Omega_{\mathrm{Th}} \eta_{-}| \to \infty$), the symmetric state produces the same dependence of the coincidence rate on the delay time $\tau$ as the asymmetric state considered in Sec.\ \ref{sec:entangled}. In contrast, the antisymmetric state ($\theta = \pi$) leads to a dip in the coincidence rate at $\tau = 0$, instead of a peak. This is indeed reminiscent of the behavior of a pair of fermions that obey the Pauli principle and therefore avoid being in the same quantum state. Instead of an increase of coincidence rate observed for the state with the bosonic symmetry, the fermionic symmetry results in a complete suppression of coincidences, making it impossible to find two photons in the same outgoing mode. This effect is suppressed when the strength of disorder is increased (see the pair of lines corresponding to $|\Omega_{\mathrm{Th}} \eta_{-}| = 0.3$ in Fig.\ \ref{fig:symantisym}), but the dip at $\tau = 0$ is still clearly visible.

\begin{figure}
\centering
\includegraphics[width=8cm]{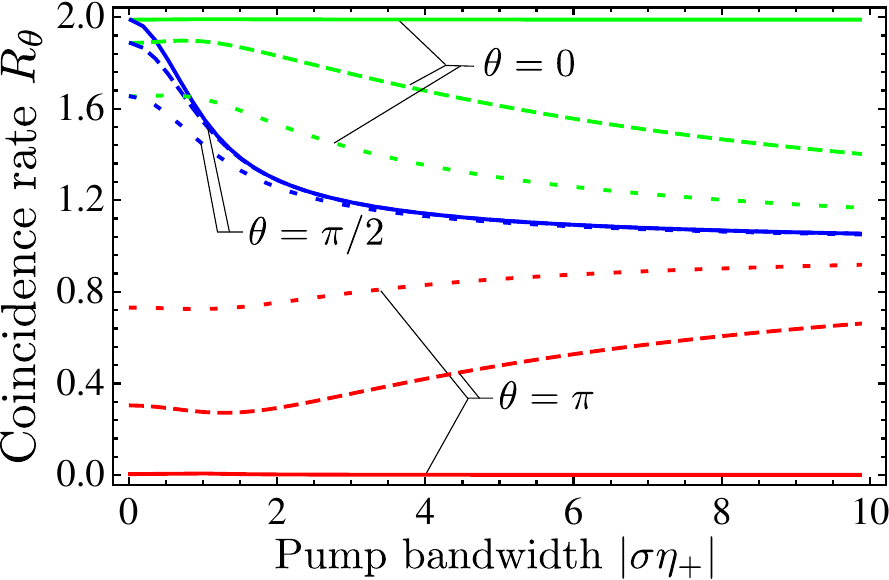}
\caption{Photocount coincidence rate in the absence of time delay ($\tau = 0$) and for $|\Omega_{\mathrm{Th}} \eta_{-}| \to \infty$ (solid lines), 1 (dashed lines) and 0.3 (dotted lines), as a function of the normalized bandwidth of the pump $|\sigma \eta_+|$.}
 \label{fig:symtau0}
\end{figure}

The role of symmetry in the structure of the entangled state becomes obvious when we look at the evolution of $R_{\theta}(\tau)$ with the bandwidth $\sigma$ of the pump. According to Fig.\ \ref{fig:spectra}, increasing $\sigma$ makes the spectra of the two photons different and thus makes the photons distinguishable. This suppresses two-photon interference effects and reduces the peak in $R_{\mathrm{ent}}(\tau)$ (see Fig.\ \ref{fig:hom}). However, the situation is quite different for the symmetrized states that we study in the present section: in the states corresponding to $\theta = 0$ or $\pi$, the two photons remain indistinguishable independent of $\sigma$. The dependence of $R_{\theta}(\tau=0)$ on $|\sigma \eta_+|$ shown in Fig.\ \ref{fig:symtau0} illustrates this quite convincingly. At weak disorder ($|\Omega_{\mathrm{Th}} \eta_{-}| \to \infty$), $R_{\theta}(\tau=0)$ does not vary with $\sigma$, whereas at stronger disorder, increasing the bandwidth eventually suppresses interferences, though at considerably larger scales as compared to the asymmetric $\theta = \pi/2$ state.

\begin{figure}
  \centering
  \includegraphics[width=8cm]{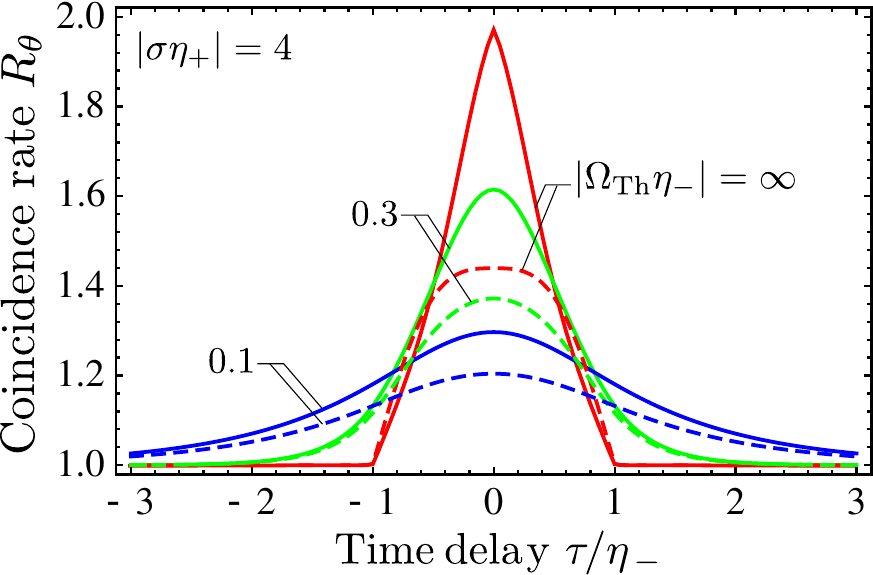}
  \caption{Comparison of photocount rates for states corresponding to $\theta = 0$ (symmetric state, solid lines) and $\pi/2$ (dashed lines), for a particular value of pump bandwidth $|\sigma \eta_+| = 4$.}
  \label{fig:pi2vs0}
\end{figure}

\begin{figure}
\centering
\includegraphics[width=8cm]{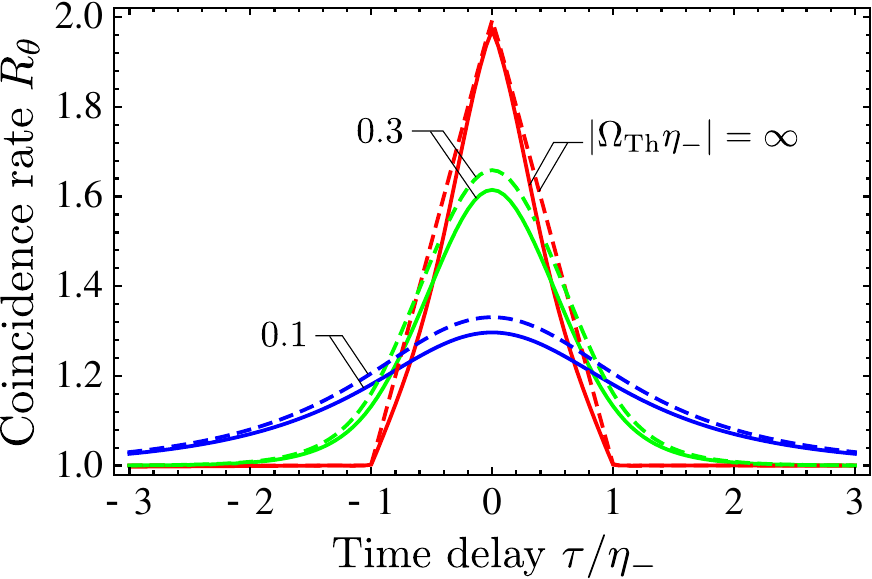}
\caption{Comparison of photocount coincidence rates corresponding to $\theta = 0$, $|\sigma \eta_+| = 4$ (solid lines) and $\theta = \pi/2$, $|\sigma \eta_+| = 0$ (dashed lines).}
\label{fig:pi2vs02}
\end{figure}

To illustrate the positive impact of the symmetry of the state on the two-photon interference, in Fig.\ \ref{fig:pi2vs0} we show the dependence of the photocount coincidence rate on the delay time for the symmetric state ($\theta = 0$) and the asymmetric state corresponding to $\theta = \pi/2$, at three different disorder strengths. For weak disorder ($|\Omega_{\mathrm{Th}} \eta_{-}| \to \infty$), the difference is very important, although at stronger disorder the two results start to be closer, showing that disorder tends to reduce the role of symmetry. To quantify the impact of symmetry even further, in Fig.\ \ref{fig:pi2vs02} we replot the solid lines corresponding to $\theta = 0$ and $|\sigma \eta_+| = 4$ and compare them with the result corresponding to the $\theta = \pi/2$ state at $|\sigma \eta_+| = 0$ (monochromatic pump). The curves of each of the 3 pairs, corresponding to different disorder strengths, are very close to each other. This illustrates that the symmetry of the state can compensate for the loss of two-photon interference due to the large bandwidth $\sigma$ of the pump pulse.

\section{Signs of nonclassical light}
\label{sec:signs}

Let us now clarify which properties of the coincidence rate $R$ measured in transmission through a disordered medium result from the classical or nonclassical nature of incident light. We adopt the following operational definition \cite{Klyshko96a,Klyshko96b}: we call light nonclassical if its photocount statistics $p(n)$ cannot be obtained from the semiclassical Mandel's formula \cite{Mandel95}
\begin{eqnarray}
p(n) = \int_0^{\infty} \frac{I^n}{n!} \exp(-I) {\cal P}(I) dI
\label{eq:mandel}
\end{eqnarray}
with ${\cal P}(I) \geq 0$ --- the probability density of the classical variable (intensity) $I$.
Obviously, the probability to detect 2 photons in a given outgoing mode $i$ readily follows: $p(2) = \overline{I^2 \exp(-I)}/2 \simeq \overline{I^2}/2$. As before, the vertical line denotes averaging over realizations of disorder that in the present context is equivalent to averaging over the distribution ${\cal P}(I)$ and we made use of the fact that in transmission through a thick disordered medium, ${\cal P}(I)$ is appreciable only for $I \ll 1$. We see that Eq.\ (\ref{eq:mandel}) implies that the normalized photocount coincidence rate as defined by Eq.\ (\ref{eq:rater}), is given by
\begin{eqnarray}
R = \frac{p(2)}{\overline{T}^2} = 2 \left( 1 +
\overline{\delta I^2}/\overline{I}^2 \right),
\label{eq:rsemiclass}
\end{eqnarray}
where $\delta I = I - \overline{I}$ is the fluctuation of intensity and we used the fact that $\overline{I} = 2 \overline{T}$ for the states considered in this work [see Eq.\ (\ref{eq:intensity1})].

In fact, we already used Eq.\ (\ref{eq:rsemiclass}) to discuss the results obtained for the coherent state in Sec.\ \ref{sec:coherent} with $I = I_{\Delta T}$. $R = 2$ was obtained in the absence of intensity fluctuations ($\overline{\delta I^2} = 0$) and $R = 4$ --- for $\overline{\delta I^2}/(\overline{I})^2 = 1$. We thus immediately conclude that the results for the coherent state can be described by Eq.\ (\ref{eq:rsemiclass}) following from the Mandel's formula (\ref{eq:mandel}), confirming that, not surprisingly, this state remains classical after transmission through the medium. In contrast, we saw in Secs. \ref{sec:entangled}, \ref{sec:separable} and \ref{sec:symmetry} that two-photon entangled and separable states lead to $R < 2$. This would require $\overline{\delta I^2}/(\overline{I})^2 < 0$ in Eq.\ (\ref{eq:rsemiclass}), which is impossible for any probability distribution ${\cal P}(I)$. Therefore, the results for two-photon states cannot be described by Eq.\ (\ref{eq:mandel}). This shows that these states remain nonclassical upon transmission through a disordered medium.

Measurement of the absolute value of $R$ in an experiment may be complicated because it requires proper normalization of the photocount number which is the raw output of the measuring device. The difference between classical and nonclassical states of light also shows up in the contrast (or visibility) of the coincidence curve $R(\tau)$:
\begin{eqnarray}
V = \frac{|R(0) - R(\infty)|}{R(0) + R( \infty)}.
\label{eq:vis}
\end{eqnarray}
This quantity is not sensitive to the normalization of $R$ and may be easier to access experimentally. The maximum contrast is reached in the limit of $\Omega_{\mathrm{Th}} \to \infty$. For the asymmetric entangled state (\ref{eq:entangledstate}), the symmetric entangled state [Eq.\ (\ref{eq:btheta}) with $\theta = 0$], and the Fock two-photon state, the maximum contrast is $\max(V_{\mathrm{ent}}) = \max(V_{\theta = 0}) = \max(V_{\mathrm{Fock}}) = 1/3$, whereas for the coherent state we find $\max(V_{\mathrm{coh}}) = 1/7$. The largest contrast $\max(V_{\theta = \pi}) = 1$ is reached for the antisymmetric entangled state [Eq.\ (\ref{eq:btheta}) with $\theta = \pi$]. Therefore, the contrast of the two-photon speckle pattern is more than a factor of 2 larger for the nonclassical light considered in this paper than for the classical light, represented by the coherent state.

\section{Conclusion}
\label{sec:conclusion}

The purpose of this paper was to discuss quantum and classical aspects of two-photon interference behind a disordered medium. We consider a particular realistic example of a light pulse of finite bandwidth sent through a disordered medium and containing 2 orthogonally polarized photons, of which one can be delayed by an arbitrary time $\tau$. By comparing the rates $R(\tau)$ of coincident photon counting in the same outgoing mode in transmission through the disordered medium for different quantum states of the incident light (two-photon entangled and Fock states, coherent state), we convincingly demonstrate that disorder is the main factor that shapes the curve $R(\tau)$. Moreover, if no additional information about the incident pulse is available, it is barely possible to see any sign of entanglement in $R(\tau)$ for a typical two-photon entangled state generated by the collinear type-II SPDC. Almost identical $R(\tau)$ curves can be obtained for an entangled and separable (Fock) two-photon states, provided that the bandwidth of the latter is adjusted. However, the result obtained for the coherent state is different, which highlights the second important factor that plays a role in the two-photon interference: the quantum nature of light. Here it is manifest in the fact that the number of photons is well-defined in the two-photon entangled and Fock states, whereas it is not a good quantum number in the coherent state. Finally, the third important aspect is entanglement that allows one to control the symmetry of the two-photon state. Symmetric and antisymmetric states that can be prepared in this way lead to coincidence rates $R(\tau)$ that do not decrease with the pump bandwidth as fast as for the standard SPDC-generated entangled state. In addition, the antisymmetric state allows one to model fermionic behavior and change the constructive two-photon interference into a destructive one. As a result, the peak of $R(\tau)$ observed for the state with the bosonic symmetry (i.e., for the symmetric state), as well as for the states with no particular symmetry, at $\tau = 0$, turns into a deep for the state with the fermionic symmetry (i.e., for the antisymmetric state).

\section*{Acknowledgements}
We acknowledge discussions with Vladimir Fedorov at the early stages of this work.

\appendix

\section{Photon number correlation versus coincidence rate}
\label{app:corrvsrate}

In this Appendix, we establish a relation between the photon number correlation function $C_{ij} = \langle {\hat n}_i {\hat n}_j \rangle$ and the probability $P_2(i,j)$ to detect a photon in each of the modes $i$ and $j$. Let us first assume $i \ne j$. Denote the state of the field {\it before\/} detection of photons by $\ket{\psi}$ and assume that $\{ \ket{\psi_n} \}$ is an orthonormal basis composed of all possible, orthogonal states in which the field can be found {\it after\/} the two photons are detected. The probability density to detect a photon from the mode $i$ at a time $t_1$ and a photon from the mode $j$ afterwards, at a time $t_2 > t_1$, is then \cite{Mandel95}
\begin{eqnarray}
\label{eq:p2ij}
P_2(i,j; t_1, t_2) &=& \sum_n
\left| \bra{\psi_n} {\hat a}_j(t_2) {\hat a}_i(t_1) \ket{\psi} \right|^2
\nonumber \\
&=& \sum_n
\bra{\psi} {\hat a}_i^{\dagger}(t_1) {\hat a}_j^{\dagger}(t_2) \ket{\psi_n} \bra{\psi_n} {\hat a}_j(t_2) {\hat a}_i(t_1) \ket{\psi}
\nonumber \\
&=& \bra{\psi} {\hat a}_i^{\dagger}(t_1) {\hat a}_j^{\dagger}(t_2) {\hat a}_j(t_2) {\hat a}_i(t_1) \ket{\psi}
\nonumber \\
&=& \langle: {\hat n}_i(t_1) {\hat n}_j(t_2) :\rangle,
\end{eqnarray}
where the operator ${\hat a}_i(t)$ corresponds to the detection of a photon in the mode $i$ at a time $t$ \cite{Mandel95} and we made use of the closure relation $\sum_n \ket{\psi_n} \bra{\psi_n} = 1$; the colons $:\cdots:$ denote normal ordering of operators.

The probability that the two photons are detected at some times $t_1 < t_2$ during a sampling time $\Delta T$ is obtained by integrating Eq.\ (\ref{eq:p2ij}) over times:
\begin{eqnarray}
\label{eq:p2ija}
P_2^{t_1 < t_2}(i,j) &=&
\int\limits_{-\Delta T/2}^{\Delta T/2} dt_1
\int\limits_{t_1}^{\Delta T/2} dt_2 P_2(i,j; t_1,t_2).
\end{eqnarray}
Finally, the probability $P_2(i,j)$ of detecting the two photons in arbitrary order is equal to the sum of $P_2^{t_1 < t_2}(i,j)$ given by Eq.\ (\ref{eq:p2ija}) and $P_2^{t_1 > t_2}(i,j)$ given by Eq.\ (\ref{eq:p2ija}) with the integration over $t_2$ running from $-\Delta T/2$ to $t_1$:
\begin{eqnarray}
\label{eq:p2ijb}
P_2(i,j) &=& P_2^{t_1 < t_2}(i,j) + P_2^{t_1 > t_2}(i,j)
\nonumber \\
&=&\int\limits_{-\Delta T/2}^{\Delta T/2} dt_1
\int\limits_{t_1}^{\Delta T/2} dt_2 P_2(i,j; t_1,t_2)
\nonumber \\
&+& \int\limits_{-\Delta T/2}^{\Delta T/2} dt_1
\int\limits_{-\Delta T/2}^{t_1} dt_2 P_2(i,j; t_1,t_2)
\nonumber \\
&=& \int\limits_{-\Delta T/2}^{\Delta T/2} dt_1
\int\limits_{-\Delta T/2}^{\Delta T/2} dt_2 P_2(i,j; t_1,t_2)
\nonumber \\
&=& \langle: {\hat n}_i {\hat n}_j :\rangle
= \langle {\hat n}_i {\hat n}_j \rangle.
\end{eqnarray}
We therefore conclude that for $i \ne j$, $P_2(i,j)$ and $C_{ij}$ are exactly equal.

Assume now that $i = j$. We have
\begin{eqnarray}
\label{eq:p2ii}
P_2(i,i; t_1, t_2)
&=& \langle: {\hat n}_i(t_1) {\hat n}_i(t_2) :\rangle
\end{eqnarray}
and
\begin{eqnarray}
\label{eq:p2iia}
P_2(i,i)
&=& \int\limits_{-\Delta T/2}^{\Delta T/2} dt_1
\int\limits_{t_1}^{\Delta T/2} dt_2 P_2(i,i; t_1,t_2)
\nonumber \\
&=& \frac12 \langle: {\hat n}_i^2 :\rangle
= \frac12 \left( \langle {\hat n}_i^2 \rangle - \langle {\hat n}_i \rangle \right).
\end{eqnarray}
The difference with respect to the case $i \ne j$ comes from the fact that the two photons now belong to the same mode and hence are indistinguishable. The two cases $t_1 < t_2$ and $t_1 > t_2$ cannot be distinguished anymore and there is only one [instead of two in Eq.\ (\ref{eq:p2ijb})] contribution to $P_2$. Equations (\ref{eq:p2ijb}) and (\ref{eq:p2iia}) lead to Eq.\ (\ref{eq:ratecorr}) of the main text.

The importance of the additional prefactor $\frac12$ in Eq.\ (\ref{eq:p2iia}) as compared to Eq.\ (\ref{eq:p2ijb}) can be understood if we consider two identical one-photon wave packets incident at the same input channel of a symmetric beam splitter having two outgoing modes $i, j = 1$ (transmission) or 2 (reflection). Assume that the wave packets are well separated in time so that they interact with the beam splitter independently and each photon can be transmitted or reflected with a probability $P_1(1) = P_1(2) = \frac12$. On the one hand, the calculation of joint probabilities readily yields $P_2(1,1) = P_2(2,2) = \frac14$ and $P_2(1,2) = \frac12$. On the other hand, we find $\langle: {\hat n}_1^2 :\rangle = \langle: {\hat n}_2^2 :\rangle = \langle: {\hat n}_1 {\hat n}_2 :\rangle = \frac12$. We thus see that although  $P_2(1,2) = \langle: {\hat n}_1 {\hat n}_2 :\rangle$, an additional factor $\frac12$ is necessary to link $P_2(1,1)$ and $\langle: {\hat n}_1^2 :\rangle$: $P_2(1,1) = \frac12 \langle: {\hat n}_1^2 :\rangle$. The difference between the cases $i = j$ and $i \ne j$ comes from the fact that two different processes can lead to detecting one photon in the mode 1 and the other --- in the mode 2: either the first photon is transmitted and the second is reflected or vice versa. The (equal) probabilities of these two processes add up to give $P_2(1,2)$. However, a unique process leads to finding both photons in the mode 1 (or 2): both photons should be transmitted (or reflected).

\section{Calculation of normally ordered photon number correlation functions}
\label{app:corrcalculation}

In this Appendix, we derive Eqs.\ (\ref{eq:coinc_ent1})--(\ref{eq:coinc_coh1}) of the main text. Consider first the case of $\tau = 0$, i.e. the case when there is no time delay between the two photon beams. We have
\begin{eqnarray}
\label{eq:corrorder}
\langle: \hat{n}_{i} \hat{n}_{j} :\rangle = \int\limits_{-\Delta T/2}^{\Delta T/2} dt_1
\int\limits_{-\Delta T/2}^{\Delta T/2} dt_2 \langle: \hat{n}_{i}(t_{1}) \hat{n}_{j}(t_{2}) :\rangle, \;\;\;\;\;\;
\end{eqnarray}
where
\begin{eqnarray}
\label{eq:corrorder2}
\langle: \hat{n}_{i}(t_{1}) \hat{n}_{j}(t_{2}) :\rangle &=&
\bra{\psi}\hat{a}^{\dagger}_{i}\left(t_{1}\right)\hat{a}^{\dagger}_{j}\left(t_{2}\right)
\hat{a}_{j}\left(t_{2}\right)\hat{a}_{i}\left(t_{1}\right)\ket{\psi}
\nonumber\\
&=& |\hat{a}_{j}\left(t_{2}\right)\hat{a}_{i}\left(t_{1}\right)\ket{\psi}|^{2}.
\end{eqnarray}
Representing ${\hat a}_i(t)$ through its Fourier transform ${\hat a}_i(\omega)$, $\hat{a}_i(t) = (1/\sqrt{2\pi})\int_{-\infty}^{+\infty}d\omega \hat{a}_i (\omega) e^{-i\omega t}$, we obtain
\begin{eqnarray}
\label{eq:corrorder3}
\hat{a}_{j}(t_{2})\hat{a}_{i}(t_{1})\ket{\psi} &=&\frac{1}{2\pi}\iint_{-\infty}^{\infty}d\omega d\omega' \hat{a}_{j}(\omega)\hat{a}_{i}(\omega')
\nonumber\\
&\times & e^{-i(\omega t_{2}+\omega' t_{1})}\ket{\psi}.
\end{eqnarray}
Using Eq.\ (\ref{eq:input-output}), we now express $\hat{a}_j (\omega)$ and $\hat{a}_i(\omega)$ through $\hat{a}_{\mathrm{o}}(\omega)$ and $\hat{a}_{\mathrm{e}}(\omega)$:
\begin{eqnarray}
\label{eq:corrorder4}
\hat{a}_{j}(t_{2}) \hat{a}_{i}(t_{1}) \ket{\psi} &=&\frac{1}{2\pi}\iint_{-\infty}^{\infty}d\omega d\omega'
\nonumber\\
&\times &\left[ t_{{\mathrm{e}}j}(\omega) \hat{a}_{{\mathrm{e}}}(\omega)+
t_{{\mathrm{o}}j}(\omega) \hat{a}_{{\mathrm{o}}}(\omega) \right]
\nonumber\\
&\times & \left[ t_{{\mathrm{e}}i}(\omega') \hat{a}_{{\mathrm{e}}}(\omega')+
t_{{\mathrm{o}}i}(\omega')
\hat{a}_{{\mathrm{o}}}(\omega') \right]
\nonumber\\
&\times & e^{-i\left(\omega t_{2}+\omega' t_{1}\right)}\ket{\psi}.
\end{eqnarray}
Note that strictly speaking, we have to keep in the expressions for $\hat{a}_j (\omega)$ and $\hat{a}_i(\omega')$ in Eq.\ (\ref{eq:corrorder4}) contributions from all the incoming modes others than the modes `o' and `e', even if no light is sent into these modes. Otherwise, the representations of the operators $\hat{a}_j (\omega)$ and $\hat{a}_i(\omega')$ corresponding to outgoing modes, in terms of operators, corresponding to incoming modes, do not verify the bosonic commutation relations. We will, however, omit these additional terms from here on because they do not contribute to the normal-ordered correlation functions that we compute in this Appendix.

For the rest of the calculation, it will be convenient to distinguish between the light in two-photon states (entangled and Fock states) and the light in the coherent state.

\subsection{Two-photon entangled and Fock states}
\label{app:entfock}

If the state $\ket{\psi}$ contains only two photons of which one has ordinary and the other one --- extraordinary polarization, we can simplify Eq.\ (\ref{eq:corrorder4}) by dropping the terms containing $\hat{a}_{{\mathrm{o}}}(\omega) \hat{a}_{{\mathrm{o}}}(\omega') \ket{\psi}$ and $\hat{a}_{{\mathrm{e}}}(\omega) \hat{a}_{{\mathrm{e}}}(\omega') \ket{\psi}$:
\begin{eqnarray}
\label{eq:corrorder5}
\hat{a}_{j}(t_{2}) \hat{a}_{i}(t_{1}) \ket{\psi} &=& \frac{1}{2\pi} \iint_{-\infty}^{\infty} d\omega d\omega'
\nonumber\\
&\times &\left[ t_{{\mathrm{e}}j}(\omega) t_{{\mathrm{o}}i}(\omega')
\hat{a}_{{\mathrm{e}}}(\omega) \hat{a}_{{\mathrm{o}}}(\omega')\right.
\nonumber\\
&+& \left. t_{{\mathrm{e}}i}(\omega')t_{{\mathrm{o}}j}(\omega)
\hat{a}_{{\mathrm{o}}}(\omega) \hat{a}_{{\mathrm{e}}}(\omega')\right]
\nonumber\\
&\times & e^{-i\left(\omega t_{2}+\omega' t_{1}\right)}\ket{\psi}.
\end{eqnarray}
Using the bosonic commutation relations obeyed by the operators $\hat{a}_{{\mathrm{e}}}(\omega)$ and $\hat{a}_{{\mathrm{o}}}(\omega')$, we obtain
\begin{eqnarray}
\label{eq:corrorder6}
\hat{a}_{{\mathrm{e}}}(\omega) \hat{a}_{{\mathrm{o}}}(\omega') \ket{\psi} &=& \iint_{-\infty}^{\infty} d\omega_{1} d\omega_{2} B(\omega_{1},\omega_{2})
\nonumber\\
&\times& \hat{a}_{{\mathrm{e}}}(\omega) \hat{a}_{{\mathrm{o}}}(\omega')
\hat{a}_{{\mathrm{o}}}^{\dagger}(\omega_{1}) \hat{a}_{{\mathrm{e}}}^{\dagger}(\omega_{2}) \ket{0}
\nonumber\\
&=& B(\omega',\omega) \ket{0},
\end{eqnarray}
and, similarly,
\begin{eqnarray}
\label{eq:corrorder7}
\hat{a}_{{\mathrm{o}}}(\omega) \hat{a}_{{\mathrm{e}}}(\omega') \ket{\psi} &=& B(\omega, \omega') \ket{0}.
\end{eqnarray}

We now assume that the sampling time $\Delta T$ during which the photons are counted is much longer than all other time scales of the problem. This allows us to take the limit $\Delta T \to \infty$ in Eq.\ (\ref{eq:corrorder}) and to use the integral relation  $\int_{-\infty}^{\infty} dt e^{-i(\omega-\Omega)t} = 2 \pi \delta(\omega-\Omega)$. We then obtain
\begin{eqnarray}
\label{eq:corrorder8}
&&\langle: \hat{n}_{i} \hat{n}_{j} :\rangle = \iint_{-\infty}^{\infty} d\omega_{1} d\omega_{2}
\left\{ |B(\omega_{1}, \omega_{2})|^{2}
\right.
\nonumber \\
&&\times \left. \left[ |t_{{\mathrm{e}}j}(\omega_{2})|^{2}
|t_{{\mathrm{o}}i}(\omega_{1})|^{2} + |t_{{\mathrm{o}}j}(\omega_{2})| ^{2}
|t_{{\mathrm{e}}i}(\omega_{1})|^{2} \right]
\right.
\nonumber\\
&&+ \left. B(\omega_{1}, \omega_{2}) B^{\ast}(\omega_{2}, \omega_{1})
\right.
\nonumber\\
&&\times \left. \left[ t_{{\mathrm{e}}j}(\omega_{2}) t_{{\mathrm{e}}i}^{\ast}(\omega_{1})
t_{{\mathrm{o}}i}(\omega_{1}) t_{{\mathrm{o}}j}^{\ast}(\omega_{2})
\right. \right.
\nonumber \\
&&+ \left. \left.
t_{{\mathrm{e}}i}(\omega_{2}) t_{{\mathrm{e}}j}^{\ast}(\omega_{1})
t_{{\mathrm{o}}j}(\omega_{1}) t_{{\mathrm{o}}i}^{\ast}(\omega_{2})
\right] \right\}.
\end{eqnarray}

We now average Eq.\ (\ref{eq:corrorder8}) over disorder by using the Gaussian statistics of transmission coefficients $t_{\alpha i}(\omega)$, as discussed in Sec.\ \ref{sec:disorder}. This yields
\begin{eqnarray}
\label{eq:corrorder9}
\overline{\langle: \hat{n}_{i} \hat{n}_{j} :\rangle}_{\mathrm{ent}} &=& 2 \overline{T}^2 \left[1 +
\delta_{ij} \iint_{-\infty}^{\infty} d\omega_{1} d\omega_{2}
B^{*}(\omega_{1},\omega_{2}) \right.
\nonumber \\
&\times& \left.  B(\omega_{2},\omega_{1})
|C(\omega_{1}-\omega_{2})|^{2}\right].
\end{eqnarray}
The result for the Fock state is obtained by replacing $B(\omega_{1},\omega_{2})$ by ${\tilde B}(\omega_{1}) {\tilde B}(\omega_{2})$:
\begin{eqnarray}
\overline{\langle: \hat{n}_{i} \hat{n}_{j} :\rangle}_{\mathrm{Fock}} &=& 2 \overline{T}^2 \left[1 +
\delta_{ij} \iint_{-\infty}^{\infty} d\omega_{1} d\omega_{2}
|{\tilde B}(\omega_{1})|^2 \right.
\nonumber \\
&\times& \left. |{\tilde B}(\omega_{2})|^2 |C(\omega_{1}-\omega_{2})|^{2}\right].
\end{eqnarray}

Equations (\ref{eq:coinc_ent1}) and (\ref{eq:coinc_sep1}) of the main text, corresponding to the situation in which the photon that has extraordinary polarization is delayed by a time $\tau$ before entering into the disordered medium, are obtained by replacing $t_{{\mathrm{e}}j}(\omega)$ by $t_{{\mathrm{e}}j}(\omega) e^{-i\omega\tau}$ in the above derivation.

\subsection{Coherent state}
\label{app:coherent}

Using the fact that the two-mode coherent state defined by Eq.\ (\ref{eq:coherentstate}) is an eigenstate of operators ${\hat a}_{\mathrm{o, e}}(\omega)$: ${\hat a}_{\mathrm{o, e}}(\omega) \ket{\psi_\mathrm{coh}} = \alpha(\omega) \ket{\psi_\mathrm{coh}}$, we reduce Eq.\ (\ref{eq:corrorder4}) to
\begin{eqnarray}
\label{eq:corrorder10}
\hat{a}_{j}(t_{2}) \hat{a}_{i}(t_{1}) \ket{\psi_{\mathrm{coh}}} &=& \frac{1}{2\pi} \int_{-\infty}^{\infty} d\omega d\omega'
\nonumber\\
&\times& \left[ t_{{\mathrm{e}}j}(\omega) t_{{\mathrm{e}}i}(\omega') e^{-i(\omega+\omega')\tau}\right.
\nonumber\\
&+& t_{{\mathrm{e}}j}(\omega) t_{{\mathrm{o}}i}(\omega') e^{-i\omega\tau}
\\
&+&  t_{{\mathrm{o}}j}(\omega) t_{{\mathrm{e}}i}(\omega') e^{-i\omega'\tau}
\nonumber\\
&+& \left. t_{{\mathrm{o}}j}(\omega) t_{{\mathrm{o}}i}(\omega') \right]
\nonumber\\
&\times& \alpha(\omega) \alpha(\omega') e^{-i(\omega t_{2}+\omega' t_{1})} \ket{\psi}.\nonumber
\end{eqnarray}
In contrast to the two-photon states considered in Sec.\ \ref{app:entfock}, the terms containing  $t_{{\mathrm{e}}j}(\omega) t_{{\mathrm{e}}i}(\omega')$ and $t_{{\mathrm{o}}j}(\omega) t_{{\mathrm{o}}i}(\omega')$ do not vanish. We now write down $|\hat{a}_{j}(t_{2}) \hat{a}_{i}(t_{1}) \ket{\psi_{\mathrm{coh}}}|^{2}$, integrate it over times $t_1$ and $t_2$ in the limit $\Delta T \to \infty$, and average over disorder. This yields
\begin{eqnarray}
\label{eq:corrorder11}
&&\overline{\langle: \hat{n}_{i} \hat{n}_{j} :\rangle}_{\mathrm{coh}} = 4 \overline{T}^2
\iint_{-\infty}^{\infty} d\omega_{1} d\omega_{2}
|\alpha(\omega_{1})|^2 |\alpha(\omega_{2})|^2
\nonumber\\
&\times&
\left[ 1+\delta_{ij} |C(\omega_{1}-\omega_{2})|^{2} \cos^{2}([\omega_{1}-\omega_{2}]\tau/2) \right].
\end{eqnarray}
Finally,  Eq.\ (\ref{eq:coinc_coh1}) of the main text is obtained after making use of the condition $\int_{-\infty}^{\infty} d\omega |\alpha(\omega)|^2 = 1$ that reflects the fact that each mode of our two-mode coherent state $\ket{\psi_{\mathrm{coh}}}$ contains one photon on average.

\bibliographystyle{apsrev}
\bibliography{biblio}

\end{document}